\RequirePackage{ifpdf}
\ifpdf 
\documentclass[pdftex]{sigma}
\else
\documentclass{sigma}
\fi

\begin{document}
\def\rf#1{(\ref{eq:#1})}
\def\lab#1{\label{eq:#1}}
\def\nonu{\nonumber}
\def\br{\begin{eqnarray}}
\def\er{\end{eqnarray}}
\def\be{\begin{equation}}
\def\ee{\end{equation}}
\def\eq{\!\!\!\! &=& \!\!\!\! }
\def\ba{\be\begin{array}{c}}
\def\ea{\end{array}\ee}
\def\foot#1{\footnotemark\footnotetext{#1}}
\def\lb{\lbrack}
\def\rb{\rbrack}
\def\llangle{\left\langle}
\def\rrangle{\right\rangle}
\def\blangle{\Bigl\langle}
\def\brangle{\Bigr\rangle}
\def\llb{\left\lbrack}
\def\rrb{\right\rbrack}
\def\Blb{\Bigl\lbrack}
\def\Brb{\Bigr\rbrack}
\def\lcurl{\left\{}
\def\rcurl{\right\}}
\def\({\left(}
\def\){\right)}
\def\v{\vert}                     
\def\bv{\bigm\vert}               
\def\lskip{\vskip\baselineskip\vskip-\parskip\noindent}
\def\mskp{\par\vskip 0.3cm \par\noindent}
\def\sskp{\par\vskip 0.15cm \par\noindent}
\def\bc{\begin{center}}
\def\ec{\end{center}}

\def\tr{\mathop{\rm tr}}                  
\def\Tr{\mathop{\rm Tr}}                  
\makeatletter
\newcommand{\rd}{\@ifnextchar^{\DIfF}{\DIfF^{}}}
\def\DIfF^#1{%
   \mathop{\mathrm{\mathstrut d}}%
   \nolimits^{#1}\gobblespace}
\def\gobblespace{\futurelet\diffarg\opspace}
\def\opspace{%
   \let\DiffSpace\!%
   \ifx\diffarg(%
   \let\DiffSpace\relax
   \else
   \ifx\diffarg[%
   \let\DiffSpace\relax
   \else
   \ifx\diffarg\{%
   \let\DiffSpace\relax
   \fi\fi\fi\DiffSpace}
\newcommand{\deriv}[3][]{\frac{\rd^{#1}#2}{\rd #3^{#1}}}
\providecommand*{\dder}[3][]{%
\frac{\rd^{#1}#2}{\rd #3^{#1}}}
\providecommand*{\pder}[3][]{%
\frac{\partial^{#1}#2}{\partial #3^{#1}}}
\newcommand{\renewoperator}[3]{\renewcommand*{#1}{\mathop{#2}#3}}
\renewoperator{\Re}{\mathrm{Re}}{\nolimits}
\renewoperator{\Im}{\mathrm{Im}}{\nolimits}
\providecommand*{\iu}%
{\ensuremath{\mathrm{i}\,}}
\providecommand*{\eu}%
{\ensuremath{\mathrm{e}}}
\def\a{\alpha}
\def\b{\beta}
\def\c{\chi}
\def\d{\delta}
\def\D{\Delta}
\def\eps{\epsilon}
\def\vareps{\varepsilon}
\def\g{\gamma}
\def\G{\Gamma}
\def\grad{\nabla}
\newcommand{\h}{\frac{1}{2}}
\def\l{\lambda}
\def\om{\omega}
\def\s{\sigma}
\def\O{\Omega}
\def\p{\phi}
\def\vp{\varphi}
\def\P{\Phi}
\def\pa{\partial}
\def\pr{\prime}
\def\ti{\tilde}
\def\wti{\widetilde}
\newcommand{\ft}{f_{\theta}}
\newcommand{\cA}{\mathcal{A}}
\newcommand{\cB}{\mathcal{B}}
\newcommand{\cC}{\mathcal{C}}
\newcommand{\cD}{\mathcal{D}}
\newcommand{\cE}{\mathcal{E}}
\newcommand{\cF}{\mathcal{F}}
\newcommand{\cG}{\mathcal{G}}
\newcommand{\cH}{\mathcal{H}}
\newcommand{\cI}{\mathcal{I}}
\newcommand{\cJ}{\mathcal{J}}
\newcommand{\cK}{\mathcal{K}}
\newcommand{\cL}{\mathcal{L}}
\newcommand{\cM}{\mathcal{M}}
\newcommand{\cN}{\mathcal{N}}
\newcommand{\cO}{\mathcal{O}}
\newcommand{\cP}{\mathcal{P}}
\newcommand{\cQ}{\mathcal{Q}}
\newcommand{\cR}{\mathcal{R}}
\newcommand{\cS}{\mathcal{S}}
\newcommand{\cT}{\mathcal{T}}
\newcommand{\cU}{\mathcal{U}}
\newcommand{\cV}{\mathcal{V}}
\newcommand{\cW}{\mathcal{W}}
\newcommand{\cX}{\mathcal{X}}
\newcommand{\cY}{\mathcal{Y}}
\newcommand{\cZ}{\mathcal{Z}}
\newcommand{\nit}{\noindent}
\newcommand{\bi}[1]{\bibitem{#1}}

\allowdisplaybreaks
\numberwithin{equation}{section}

\renewcommand{\PaperNumber}{070}

\FirstPageHeading

\renewcommand{\thefootnote}{$\star$}

\ShortArticleName{On a Negative Flow of the AKNS Hierarchy}

\ArticleName{On a Negative Flow of the AKNS Hierarchy\\ and Its  Relation 
to a Two-Component\\ Camassa--Holm Equation\footnote{This paper is a contribution 
to the Proceedings of the O'Raifertaigh Symposium 
on Non-Perturbative and Symmetry Methods in Field Theory
 (June 22--24, 2006, Budapest, Hungary).
The full collection is available at 
\href{http://www.emis.de/journals/SIGMA/LOR2006.html}{http://www.emis.de/journals/SIGMA/LOR2006.html}}}

\Author{Henrik ARATYN~$^\dag$ and Jose Francisco GOMES~$^\ddag$ and Abraham H.~ZIMERMAN~$^\ddag$}
\AuthorNameForHeading{H.~Aratyn, J.F.~Gomes and A.H. Zimerman}

\Address{$^\dag$~Department of Physics, University of Illinois at Chicago,\\
$\phantom{^\dag}$~845 W. Taylor St., Chicago, Illinois 60607-7059}
\EmailD{\href{mailto:aratyn@uic.edu}{aratyn@uic.edu}}

\Address{$^\ddag$~Instituto de F\'{\i}sica Te\'{o}rica-UNESP, Rua Pamplona 145,
01405-900 S\~{a}o Paulo, Brazil}
\EmailD{\href{mailto:jfg@ift.unesp.br}{jfg@ift.unesp.br}, \href{mailto:zimerman@ift.unesp.br}{zimerman@ift.unesp.br}}

\ArticleDates{Received September 13, 2006, in f\/inal form October 05,
2006; Published online October 17, 2006}

\Abstract{Dif\/ferent gauge copies of the Ablowitz--Kaup--Newell--Segur (AKNS) model labeled by
an angle $\theta$
are constructed and then reduced to the two-component
Camassa--Holm model. Only three dif\/ferent independent classes of
reductions are encountered corresponding to the angle $\theta$
being $0$, $\pi/2$ or taking any value
in the interval $0<\theta<\pi/2$.
This construction induces B\"acklund transformations
between solutions of the two-component Camassa--Holm model
associated with dif\/ferent classes of reduction.}


\Classification{37K10; 35Q53; 53A07; 53B50}

\section{Introduction}
It is widely known that the standard integrable hierarchies
can be supplemented by a set of commuting f\/lows of a negative
order in a spectral parameter \cite{Aratyn:2001pz}.
A standard example is provided by the modif\/ied KdV-hierarchy, which can be embedded
in a new extended hierarchy. This extended hierarchy contains in addition
to the original modif\/ied KdV equation
also the dif\/ferential equation of the  sine-Gordon model
realized as the f\/irst negative f\/low \cite{kdv-sg1,kdv-sg2,kdv-sg3,kdv-sg4,kdv-sg5,kdv-sg6}.

Quite often the negative f\/lows can only be realized in a form of
non-local integral dif\/ferential equations. The cases where the negative f\/low
can be cast in form of local dif\/ferential equation which has physical
application are therefore of special interest.
Recently in \cite{CH2}, a negative f\/low
of the extended  AKNS hierarchy \cite{Aratyn:2000wr} was identif\/ied
with a two-component generalization of the Camassa--Holm equation.
The standard Camassa--Holm equation \cite{ch1,fokas}
\be
u_t -u_{txx}= - 3u u_x + 2 u_x u_{xx}  + u u_{xxx} -\kappa u_x,\qquad
\kappa={\rm const}
\lab{ch1}
\ee
enjoys a long history of serving as
a model of long waves in shallow water.
The two-component extension \cite{CH2,falqui} dif\/fers from
equation \rf{ch1} by presence on the right hand side
of a new term $\rho \rho_x$, with the new variable $\rho $ obeying
the continuity equation $\rho_t+ (u\rho)_x=0$.
Such generalization was f\/irst encountered in a study of
deformations of the bihamiltonian structure of hydrodynamic type~\cite{LiuZhang}.
Various multi-component generalizations of the
Camassa--Holm model have been subject of intense
investigations in recent literature \cite{sh4,sh5,ivan6,sh8,sh9}.

A particular connection between extended AKNS model and
a two-component generalization of the Camassa--Holm equation
was found in \cite{CH2} and in \cite{falqui}.
It was pointed out in \cite{Aratyn:2005pg}
that the second order spectral equation for a two-component
Camassa--Holm model can be cast in form of the f\/irst order
spectral equation which after appropriate gauge transformations
f\/its into an $sl(2)$ setup of linear spectral problem and associated
zero-curvature equations.

The goal of this article is to formulate a general scheme for
connecting an extended AKNS model  to a two-component
Camassa--Holm model which would encompass all known ways of
connecting the solution $f$  of the latter model to variables $r$
and $q$ of the former model. Our approach is built on making gauge
copies of an extended AKNS model labeled by angle $\theta$
belonging to an interval $ 0\le \theta \le \pi/2$ and then by
elimination of one of two components of the $sl(2)$ wave function
reach a second order non-linear partial dif\/ferential equation
which governs the
 two-component Camassa--Holm model.
We found that the construction naturally decomposes into three
dif\/ferent classes depending on whether angle $\theta$ belongs
to an interior of interval $ 0\le \theta \le \pi/2$ or
is equal to one of two boundary values unifying therefore the results of~\cite{CH2} and~\cite{wu}.
The map between these three cases induces a B\"{a}cklund like
transformations between dif\/ferent solutions $f$ of
the two-component Camassa--Holm equation.

\section{A simple derivation of a relation between
AKNS\\ and two-component Camassa--Holm models}

Our starting point is a standard f\/irst-order linear spectral
problem of the AKNS model:
\be
\Psi_y =   \( \l \sigma_3 +{ \cA_0} \) \Psi=\l \begin{bmatrix} 1 & 0 \\ 0& - 1 \end{bmatrix} \Psi
+ \begin{bmatrix} 0 & q \\ r& 0 \end{bmatrix}  \Psi  ,\lab{akns-p}
\ee
where $\l$ is a spectral parameter, $y$ a space variable and
$\Psi$ a two-component object:
\be \Psi = \begin{bmatrix}
\psi_1 \\ \psi_2 \end{bmatrix}.
\lab{2psi}
\ee
In addition, the system is augmented by a negative f\/low def\/ined
in terms of a matrix, which is inverse proportional to $\l$:
\be
\Psi_s = D^{(-1)} \Psi= \frac{1}{\l} \begin{bmatrix} A & B \\ C& - A
\end{bmatrix}\Psi . \lab{akns-n}
\ee
The compatibility condition arising from equations~\rf{akns-p} and
\rf{akns-n}:
\br
(\cA_0)_s - D^{(-1)}_y + \left[ \l \sigma_3 +\cA_0 , D^{(-1)} \right]=0  .
\lab{t-1}
\er
has a general solution:
\be
D^{(-1)}= \frac{1}{4 \b \l}   M_0 \sigma_3 M_0^{-1}
, \qquad \cA_0 = M_{0\,y}  M_0^{-1} ,
 \lab{lrakns}
\ee
in terms of the zero-grade group element, $M_0$, of ${\rm SL}(2)$.
Note that the solution, $D^{(-1)}$, of the compatibility
condition is connected to $(1/\l) \sigma_3$-matrix
by a similarity transformation.

The factor $1/4\b$ in \rf{lrakns} is a general proportionality
factor which implies a determinant formula:
\be
A^2+BC=\frac{1}{16\b^2}
\lab{det-formula}
\ee
for the matrix elements of $D^{(-1)}$ .

From \rf{t-1} we f\/ind that
\begin{gather*}
\big( \Tr(\cA_0^2) \big)_s = 2 \Tr(\cA_0 \cA_{0\,s} )
= - 2 \Tr\big(\cA_0 \big[ \l \sigma_3 , D^{(-1)} \big] \big)
= 2 \Tr\big(\l \sigma_3 \big[ \cA_0 , D^{(-1)} \big] \big)\\
\phantom{\big( \Tr(\cA_0^2) \big)_s}{} =2 \Tr \big(\l \sigma_3 D^{(-1)}_y\big) = 4 A_y
\end{gather*}
or
\be
A_y = \h (r q)_s  .
\lab{ayrqs}
\ee

When projected on the zero and the f\/irst powers of $\l$ the
compatibility condition  \rf{t-1} yields
\be
q_s = -2 B , \qquad r_s = 2 C   ,\lab{rsqs}
\ee
and
\be A_y =  q C - r B, \qquad B_y=-2 A q, \qquad C_y=2A r ,
\lab{abgy}
\ee
respectively.
Note that the f\/irst of equations \rf{abgy} together with equations \rf{rsqs}
reproduces formula~\rf{ayrqs}.

Combining the above equations we f\/ind that
\be
A= -\frac{B_y}{2 q} = \frac{q_{sy}}{4 q} =
\frac{C_y}{2 r} = \frac{r_{sy}}{4 r} .
\lab{alpa-eqs}
\ee
The spectral equation \rf{akns-p} reads in components:
\be
\psi_{1\,y}= \l \psi_1 +q \psi_2,\qquad
\psi_{2\,y}= -\l \psi_2 +r \psi_1 .
\lab{comp-spec}
\ee
Now we eliminate the wave-function component $\psi_2$
by substituting
\[ \psi_2= \frac{1}{q} \( \psi_{1y} - \l \psi_1\)
\]
into the remaining second equation of \rf{comp-spec}.
In this way we obtain for $\psi_1$
\[
\psi_{1yy} - \frac{q_y}{q} \psi_{1y}+ \frac{\l q_y}{q} \psi_{1}
-\l^2 \psi_1-rq \psi_1=0.
\]
Introducing
\be
\psi= e^{-\int p \rd y} \psi_1
\lab{psi-int}
\ee
with the integrating factor
\[
p (y)= \frac{1}{2} \(\ln q\)_y
\]
allows to eliminate the term with $\psi_{1y}$ and obtain
\be
\psi_{yy}=\left(\l^2-\l\(\ln q\)_y  -Q \right)\psi
\lab{psiyy}
\ee
with{\samepage
\be
Q = \frac{1}{2} \(\ln q\)_{yy} -\frac{1}{4} \(\ln q\)_{y}^2
-r q = \frac{q_{yy}}{2q} - \frac{3}{4}\(\frac{q_{y}}{q}\)^2
-rq
\lab{Q-def}
\ee
as in \cite{wu}.}

Eliminating $\psi_2 $ from equation \rf{akns-n} yields for
$\psi$ the following equation:
\be
\psi_s = \frac{1}{4\l} \(\frac{q_{s}}{q}\)_y \psi -
\frac{1}{2\l} \frac{q_{s}}{q} \psi_y .
\lab{psi-s}
\ee
Compatibility equation $\psi_{yys}-\psi_{syy}=0$ yields
\be
\( \frac{q_{sy}}{4q}\)_y = \h (r q)_s
\lab{compata}
\ee
in total agreement with \rf{ayrqs}.
To eliminate $r$ from \rf{compata}
we use that
\be r =  \frac{-A_y+q C}{B}
\lab{r-der}
\ee
as follows from the f\/irst equation from \rf{abgy}. Replacing
$C$  by $ 1/(B16\b^2)- A^2/B$
as follows from the determinant relation
\rf{det-formula} and recalling that $B =-q_s/2$  according to
equation \rf{rsqs} we obtain after substituting $r$ from
\rf{r-der} into \rf{compata}:
\be
\( \frac{q_{sy}}{q}\)_y = \( \frac{q_{syy}}{q_s} -\frac{q_{sy}q_y}{qq_s}
+ \frac{1}{2 \b^2} \frac{q^2}{q^2_s}
-\frac{q_{sy}^2}{2q^2_s} \)_s.
\lab{CH-q}
\ee
Note that alternatively we could have eliminated $q$ from
equation
\[
\( \frac{r_{sy}}{4r}\)_y = \h (r q)_s
\]
and obtained an equation for $r$ only. It turns out that the
equation for $r$ follows from equation~\rf{CH-q} by simply substituting
$r$ for $q$.

For brevity we introduce, as in \cite{wu}, $ f =\ln q$. Then
expression \rf{CH-q} becomes:
\be
\( f_s f_y\)_y = -\( \frac{f_{y}^2}{2} +\frac{f_{sy}^2}{2f_s^2}
 -\frac{1}{2\b^2 f_{s}^2} - \frac{f_{syy}}{f_s} \)_s  .
\lab{CH-f}
\ee
The above relation can be cast in an equivalent form:
\begin{equation}  \frac{f_{ss}}{2\b^2 f^3_{s}}+
f_{sy}f_y   +\frac{1}{2} f_s f_{yy} - \frac{f_{ssyy}}{2f_s}
+\frac{f_{ssy} f_{sy}}{2 f_s^2 } +\frac{f_{ss} f_{syy}}{2 f_s^2 }
-\frac{f_{ss} f_{sy}^2}{2 f_s^3 }=0,
\label{eq:bilcn0}
\end{equation}
which f\/irst appeared in \cite{CH2}.
The relation (\ref{eq:bilcn0}) is also equivalent to the following condition
\begin{equation}
\left(\frac{1}{f_s}\right)_s
=\b^2 \left(f_s^2 f_y - f_{ssy} +\frac{f_{ss}f_{sy}}{f_s} \right)_y \, .
\label{eq:bilcn1}
\end{equation}
For a quantity $u$ def\/ined as:
\begin{equation}
u = \b^2 \(f_s^2 f_y - f_{ssy} +\frac{f_{ss}f_{sy}}{f_s}\) -\frac{1}{2}
\kappa,
\label{eq:ufsy}
\end{equation}
with $\kappa$ being an integration constant, it holds from
relation (\ref{eq:bilcn1}) that
\begin{equation}
u_y = \left(\frac{1}{f_s}\right)_s .
\label{eq:uyfs}
\end{equation}
Next, as in \cite{Aratyn:2006vc}, we def\/ine a quantity $m$ as
$ \b^2 f_s^2 f_y$ and derive from relations  (\ref{eq:ufsy})
and (\ref{eq:uyfs}) that
\begin{gather}
m = \b^2 f_s^2 f_y= u +\b^2\(f_{ssy} -\frac{f_{ss}f_{sy}}{f_s}\) +
\frac{1}{2} \kappa
=u - \b^2 f_s \left(f_s \left(\frac{1}{f_s}\right)_s \right)_y+
\frac{1}{2} \kappa\nonumber\\
\phantom{m}=u -  \b^2 f_s (f_s  u_y)_y +\frac{1}{2} \kappa  .
\label{eq:mdef}
\end{gather}
Taking a derivative of $m$ with respect to $s$ yields
\begin{gather}
m_s = \b^2 \(2 f_y f_s f_{ss}+ f_s^2 f_{sy}\)=
2 m \frac{f_{ss}}{f_s} + \b^2 f_s^2 f_{sy}=
- 2 m f_s \left(\frac{1}{f_s}\right)_s + \b^2 f_s^2 f_{sy}\nonumber\\
\phantom{m_s}{}= -2 m f_s u_y + \b^2 f_s^2 f_{sy} .
\label{eq:mseq}
\end{gather}
In terms of quantities
$u$ and $\rho=f_s$
equations (\ref{eq:uyfs}) and (\ref{eq:mseq}) take the
following form
\begin{gather}
\rho_s = -\rho^2 u_y, \label{eq:conta}\\
m_s = -2 m \rho u_y + \b^2 \rho^2 \rho_{y} ,
\label{eq:ch1a}
\end{gather}
for $m$ given by
\begin{equation}
m= u -  \b^2 \rho  (\rho u_y)_y +\frac{1}{2} \kappa .
\label{eq:mdef1}
\end{equation}
An inverse reciprocal transformation
$(y,s)\mapsto (x,t) $ is def\/ined by relations:
\begin{equation}\label{eq:reciprocal-a}
F_x=\rho F_y ,\qquad
F_t= F_s-\rho u F_y
\end{equation}
for an arbitrary function $F$.
Equations (\ref{eq:conta}), (\ref{eq:ch1a}) and (\ref{eq:mdef1})
take a form
\begin{gather}
\rho_t  = - \left( u \rho \right)_x, \label{eq:contb}\\
m_t= -2 m u_x -m_x u+ \b^2 \rho \rho_x, \label{eq:ch2}\\
m =u - \b^2 u_{xx} + \frac{1}{2} \kappa   \label{eq:m-def}
\end{gather}
in terms of the $ (x,t) $ variables.
Equation (\ref{eq:contb}) is called the compatibility condition,
while equa\-tion~(\ref{eq:ch2}) is the two-component
Camassa--Holm equation \cite{CH2}, which agrees with
standard Camassa--Holm equation \rf{ch1} for $\rho=0$.

\section{General reduction scheme from AKNS system\\ to the two-component
Camassa--Holm equation}

Next, we perform the transformation
\be
\Psi \; \to \; \cU (\theta, f) \Psi = \begin{bmatrix} \vp\\ \eta
\end{bmatrix}
\lab{psitop}
\ee
on  AKNS two-component $\Psi$ function from
\rf{2psi}.
 $\cU (\theta, f)$ stands for an orthogonal matrix:
\be
\cU (\theta, f) = \Omega (\theta)
 \exp \( - \h f  \sigma_3
\), \qquad 0 \le \theta \le \frac{\pi}{2} ,
\lab{defU}
\ee
where $\Omega (\theta)$ is given by{\samepage
\be
 \Omega (\theta) = \sigma_3 e^{\iu \theta \sigma_2} =
\begin{bmatrix} \cos \theta & \sin \theta \\ \sin \theta & -\cos \theta
\end{bmatrix}
\lab{Omdef}
\ee
and $f$ is a function of $y$ and $s$, which is going to be
determined below for each value of $\theta$.}

Note that $\Omega^{-1} (\theta)=\Omega (\theta)$ and
$\Omega (0)=\sigma_3$, $\Omega (\pi/2)=\sigma_1$.

Taking a derivative with respect to $y$ and $s$ on both sides of
\rf{psitop} one gets
\begin{gather}
\begin{bmatrix}
\vp\\ \eta \end{bmatrix}_y = \( \cU_y \cU^{-1}+ \cU
\begin{bmatrix} \l & q \\ r& - \l \end{bmatrix} \cU^{-1} \)
\begin{bmatrix}
\vp\\ \eta \end{bmatrix},
\lab{vpy}\\
\begin{bmatrix}
\vp\\ \eta \end{bmatrix}_s = \biggl( \cU_s \cU^{-1}+ \cU
D^{(-1)} \cU^{-1} \biggr)
\begin{bmatrix}
\vp\\ \eta \end{bmatrix} . \lab{vps}
\end{gather}
Thus, the f\/lows of the new two-component function def\/ined
in \rf{defU} are governed by the gauge transformations
of the AKNS matrices $\l \sigma_3+\cA_0$ and $D^{(-1)}$, respectively.
This ensures that the original AKNS compatibility condition
\rf{t-1} still holds for the rotated system def\/ined by equations~\rf{vpy} and \rf{vps}.

{}From equation \rf{vpy} we derive that:
\begin{gather}
\l  \( \vp \cos (2 \theta) +\eta \sin (2 \theta)  \)
= \vp_y +\h \vp \(f_y \cos (2 \theta)
- \sin (2 \theta) \( q e^{-f}   + r  e^{f}\) \)\nonumber\\
\phantom{\l  \( \vp \cos (2 \theta) +\eta \sin (2 \theta)  \) =}{}
+ \eta \(\h f_y \sin (2 \theta) -r  e^{f} \sin^2  ( \theta)
+  q e^{-f} \cos^2  ( \theta) \).
\lab{vpfsy}
\end{gather}

Repeating derivation with respect to $y$ one more time yields
\begin{gather}
\begin{bmatrix}
\vp\\ \eta \end{bmatrix}_{yy} =
\biggl[ \( \cU_y \cU^{-1}+ \cU
\begin{bmatrix} \l & q \\ r& - \l \end{bmatrix} \cU^{-1} \)_y
+\( \cU_y \cU^{-1}+ \cU
\begin{bmatrix} \l & q \\ r& - \l \end{bmatrix} \cU^{-1} \)^2
\biggr] \begin{bmatrix} \vp\\ \eta \end{bmatrix}  \nonumber\\
\phantom{\begin{bmatrix} \vp\\ \eta \end{bmatrix}_{yy}}{} = \cU \begin{bmatrix} \l^2-\l f_y + f_y^2/4 -f_{yy}/2 + qr  &
q_y -f_y q  \\ r_y+f_y r & \l^2-\l f_y + f_y^2/4 +f_{yy}/2 + qr
\end{bmatrix} \cU^{-1}
\begin{bmatrix} \vp\\ \eta \end{bmatrix}.\!\!\!
\lab{vpetayy}
\end{gather}
For
\[
\begin{bmatrix}
{\bar \vp}\\ {\bar \eta} \end{bmatrix}=\Omega (\theta)\, \begin{bmatrix}
\vp\\ \eta \end{bmatrix}
\]
the result is
\begin{gather*}
\begin{bmatrix}
{\bar \vp}\\ {\bar \eta} \end{bmatrix}_{yy} =
\begin{bmatrix} \l^2-\l f_y + f_y^2/4 -f_{yy}/2 + qr  &
(q_y -f_y q)e^{-f}  \\ (r_y+f_y r)e^f & \l^2-\l f_y + f_y^2/4 +f_{yy}/2 + qr
\end{bmatrix}
\begin{bmatrix}{\bar \vp}\\ {\bar \eta} \end{bmatrix}
\end{gather*}
and shows in a transparent way that the condition
for eliminating ${\bar \eta}$ from the equation for
${\bar \vp}_{yy} $ requires $(q_y -f_y q) \exp (-f)=0 $ or
$ q= \exp(f)$. Similarly, the condition
for eliminating ${\bar \vp}$ from the equation for
${\bar \eta}_{yy} $ requires $(r_y+f_y r) \exp (f)=0 $ or
$ r= \exp(-f)$. Clearly these reductions reproduce results of the previous
section.

To obtain a more general result we return to equation \rf{vpetayy}.
Projecting on the  $\vp$-component in equation \rf{vpetayy} gives
\begin{gather}
\vp_{yy} =  \l^2 \vp - \l f_y  \vp  + \(\frac{1}{4} f_y^2 +qr\)\vp
+ \( -\h f_{y} \cos ( 2 \theta)  + \h ( q e^{-f}
  + r  e^{f}) \sin (2 \theta) \)_{\!\! y} \vp \nonumber\\
\phantom{\vp_{yy} = }{}+\(-\h f_{y}\sin (2 \theta)- q e^{-f} \cos^2 \theta
+r e^{f}  \sin^2 \theta \)_{\!\! y} \eta.
\lab{p-comp}
\end{gather}
Next, we will eliminate $\eta$ in order to obtain an equation for
the one-component variable $\vp$. This is analogous
to the calculation made below equation \rf{comp-spec}, where
the  f\/irst order two-component AKNS spectral problem
was reduced to second order equation for
the one-component function~$\psi$.
To accomplish the task we must choose $f$ so that the identity
\be
\h f_y \sin (2 \theta)= r e^f  \sin^2 \theta- q e^{-f} \cos^2 \theta+c_0
\lab{fidentity}
\ee
holds, where $c_0$ is an integration constant.
The identity \rf{fidentity} ensures that terms with $\eta$ drop out of
equation \rf{p-comp}.

Note, that for $\theta=\pi/4$ and $c_0=0$ we recover identity
$f_y = r \exp(f) - q \exp (-f)$ from \cite{CH2,Aratyn:2005pg}.
For $\theta=0$, $c_0=1$ and $\theta=\pi/2$, $c_0=-1$ we get, respectively,
$q=\exp(f)$ and $r=\exp(-f)$ as in~\cite{wu}.
From now on we take $c_0=0$ as long as $0 < \theta <\pi/2$.

Let us shift a function $f$ by a constant term,
$\ln \( \tan \theta\)$:
\be
f \longrightarrow \ft = f +\ln \( \tan \theta\).
\lab{ftdef}
\ee
Then relation \rf{fidentity} can be rewritten for $0 < \theta <\pi/2$
as
\be
f_{\theta y} = r e^{\ft} - q e^{-\ft}
\lab{ftidentity}
\ee
which is of the same form as the relation found in reference
\cite{CH2}.
It therefore appears that for all values of $\theta$ in the
the $0 < \theta <\pi/2$ relation between function $f$ and AKNS
variables~$q$ and~$r$ remains invariant up to shift of
$f$ by a constant.

Now, we turn our attention back to equation \rf{vps} rewritten as
\[
\begin{bmatrix}
\vp\\ \eta \end{bmatrix}_s =
\cU \(-\h f_s  \sigma_3 + \frac{1}{\l}
\begin{bmatrix} A & B \\ C& - A \end{bmatrix}\) \cU^{-1}
\begin{bmatrix}
\vp\\ \eta \end{bmatrix}  .
\]
For the $\vp$ component we f\/ind:
\begin{gather}
\vp_s = -\h f_s  \( \vp \cos (2 \theta) +\eta \sin (2 \theta)  \)
+ \frac{1}{2\l}\vp \(  2 A \cos (2 \theta) +
C e^f\sin (2 \theta) +Be^{-f}\sin (2 \theta)    \right)\nonumber\\
\phantom{\vp_s =}{}+\frac{1}{\l} \eta \left(
A \sin (2 \theta) +
C e^f\sin^2 \theta-Be^{-f}\cos^2 \theta \right).
\lab{vps1}
\end{gather}
For $0 < \theta < \pi/2$ we choose
\begin{equation}
B=  \left( A-\frac{1}{4\b} \right)  e^{\ft} ,
\qquad C= -
\left( A+\frac{1}{4\b}\right) e^{-\ft}  ,
\label{eq:bcf}
\end{equation}
which agrees with the determinant formula $A^2+BC=1/16\b^2$
and implies identities:
\begin{gather}
2 A - B e^{-\ft}  + C e^{\ft} =0,
\lab{asin}\\
 B e^{-\ft}+C e^{\ft} =- \frac{1}{2\b} .
\lab{acos}
\end{gather}
The f\/irst of these identities, \rf{asin},
ensures that the last three terms containing
$\eta$ on the right hand side of equation
\rf{vps1} cancel.

Recall at this point relation \rf{vpfsy}.
Simplifying this relation by invoking identity \rf{fidentity}
and plugging it into equation \rf{vps1} gives
\begin{gather}
\vp_s = -\frac{f_s}{2 \l} \vp_y
+ \frac{1}{\l} \vp \( -\frac{1}{4} f_sf_y  \cos (2 \theta)
+ \frac{f_s}{4}\sin (2 \theta) \( q e^{-f}   + r  e^{f}\) \right.\nonumber\\
\left. \phantom{\vp_s =}{}+A \cos (2 \theta) + \h B e^{-f} \sin (2 \theta) + \h
C e^f \sin (2 \theta) \).
\lab{vps2}
\end{gather}
From \rf{bcf} we f\/ind
\begin{gather}
r e^{\ft} = \frac{C_y}{2A} e^{\ft} = \frac{1}{2A}
\left(f_y  \left( A+\frac{1}{4\b} \right) -A_y \),
\lab{ref}
\\
q e^{-\ft}= -\frac{B_y}{2A} e^{-\ft} =
  \frac{-1}{2A}
\left( f_y  \left( A-\frac{1}{4\b} \right) +A_y \right)
\lab{qef1}
\end{gather}
and therefore
\be
 q e^{-\ft}   + r  e^{\ft} =
 \frac{f_y}{4A\b} -\frac{A_y}{A}.
\lab{qefref}
\ee
Due to the above relation and identity \rf{acos}
equation \rf{vps2} becomes
\be
\vp_s = -\frac{f_s}{2 \l} \vp_y
+ \frac{1}{\l} \vp \( -\frac{1}{4\b} \( 1-\frac{f_sf_y}{4 A}\)
 -\frac{A_y}{A} \frac{f_s}{4}\).
\lab{vpss2}
\ee
Taking derivative of \rf{fidentity} with respect to $s$
we f\/ind
\be
\h f_{sy}= C e^{\ft} + B e^{-\ft}
+ \h f_s\( q e^{-\ft}    + r  e^{\ft}\)
=-\frac{1}{2\b}+\frac{f_sf_y}{8 A \,\b} -\frac{f_s A_y}{2A}.
\lab{hfsy}
\ee
Thus equation \rf{vpss2} becomes
\be
\vp_s = -\frac{f_s}{2 \l} \vp_y
+ \frac{f_{sy} }{4\l} \vp   .
\lab{vps3}
\ee
We now turn our attention to equation \rf{p-comp}. The last term
containing $\eta$ vanishes due to the identity \rf{fidentity}.
In addition it holds that
\be
\frac{f_{sy}}{2 f_s} +\frac{1}{2 \b f_s}
= -\h f_{y} \cos ( 2 \theta)+\h \( q e^{-f}   + r  e^{f}\) \sin (2 \theta)
=\h\( q e^{-\ft}
  + r  e^{\ft}  \)
\lab{fid2}
\ee
as follows from relations \rf{qefref} and \rf{hfsy}.
Also, it holds from relations \rf{ref}--\rf{qef1} that
for $0 < \theta < \pi/2$:
\be
rq= \(\frac{f_{sy}}{2 f_s} +\frac{1}{2 \b f_s}\)^2 - \frac{1}{4} f_y^2
= g^2-f_y^2/4 ,
\lab{rqf}
\ee
where
\be
g= \frac{f_{sy}}{2 f_s} +\frac{1}{2 \b f_s}.
\lab{gdef}
\ee
Thus, the remaining constant  (the ones which do not
contain $\l$) terms  on the right hand side of equation \rf{p-comp}
are equal to
\begin{gather}
\frac{1}{4} f_y^2 +qr+\( -\h f_{y} \cos ( 2 \theta)  + \h ( q e^{-f}
  + r  e^{f}) \sin (2 \theta) \)_y\nonumber\\
\qquad{}=\frac{1}{4} f_y^2 +qr+ \h\( q e^{-\ft}
  + r  e^{\ft}  \)_y
= g^2 +g_y.
\lab{qrgy}
\end{gather}
Therefore, we  can  write equation \rf{p-comp} as:
\be
\vp_{yy} =  \( \l^2  - \l f_y    - Q\) \vp, \qquad
Q= -g^2 -g_y
\lab{Q-defg}
\ee
with $g$ given by \rf{gdef}.
The above spectral problem together with equation \rf{vps3}
ensures via compatibility condition $\vp_{yys}-\vp_{syy}=0$, that
\be
Q_s + \h f_{yy} f_s + f_y f_{sy}=0
\lab{qsch}
\ee
holds. The latter is equivalent to the two-component Camassa--Holm equation
\rf{CH-f}.

\section[The $\theta=0$ case and B\"acklund transformation
between different solutions]{The $\boldsymbol{\theta=0}$ case and B\"acklund transformation\\
between dif\/ferent solutions}

We now consider $\theta$ at the boundary of the
$0 < \theta < \pi/2$ interval. For illustration we take $\theta=0$,
the remaining case $\theta=\pi/2$ can be analyzed in a similar way.
Plugging $\theta=0$ into relation \rf{qrgy} we obtain
\[
rq \v_{\theta=0} =-\frac{1}{4} f_y^2 +\h f_{yy}+g^2+g_y=
 g^2 -\frac{1}{4} f_y^2 +\(\h f_y+g\)_y.
\]
Comparing with relation \rf{rqf} we get
\be
rq \v_{\theta=0} =rq \v_{\theta}+ \(\h f_y+g\)_y
\lab{rqth0}
\ee
which describes a relation between the product $rq$ for
zero and non-zero values of the angle $\theta$, with
$rq \v_{\theta}$ being associated with $\theta$ within an interval
$0<\theta <\pi/2$.

Recall that $q=\exp (f)$ for $\theta=0$.
It follows
that $A=q_{sy}/4q= (f_{sy}+f_sf_y)/4$ and equation~\rf{ayrqs}
is equivalent to
\be
\(rq \v_{\theta=0}\)_s  = \h (f_{sy}+f_sf_y)_y.
\lab{rqzero}
\ee
On the other hand, it follows from \rf{r-der}
and $ C=1/(16\b^2 B)- A^2/B$
that
\[ rq \v_{\theta=0} = \h \( f_{yy}-\h f_y^2-
\frac{f_{sy}^2}{2 f_s^2}
+\frac{1}{2 \b^2 f_s^2}+ \frac{f_{syy}}{f_s}\)
\]
and accordingly equation \rf{rqzero} is equivalent to the two-component Camassa--Holm 
equa\-tion \rf{CH-f}.

From \rf{qef1} one f\/inds for $0<\theta <\pi/2$
that:
\be
q  = \cP_{-}  (\ft)e^{\ft} ,
\lab{qef2}
\ee
where
\[
\cP_{\pm} (f) = \pm \h f_y+g =  \pm \frac{f_y}{2} +\frac{f_{sy}}{2 f_s}
+\frac{1}{2 \b f_s}.
\]
Obviously $\cP_{\pm} (\ft)=\cP_{\pm} (f)$.

We are now ready to show that
\[
{\bar f}= \ft+\ln \(  \cP_{-} (\ft) \)=
f_{\theta} + \ln \(  -
\frac{f_{\theta\,y}}{2} +\frac{f_{\theta \, sy}}{2 f_{\theta\, s}}
+\frac{1}{2 \b f_{\theta\, s}} \)
\]
satisf\/ies the two-component Camassa--Holm equation
\rf{CH-f} for any $f$ or $\ft$, which satisf\/ies
equation \rf{CH-f}.
For $0<\theta <\pi/2$, it holds that $q = \exp ({\bar f})$
and therefore
\be
A=q_{sy}/4q= ({\bar f}_{sy}+{\bar f}_s {\bar f}_y)/4=
(f_{sy}+f_sf_y)/4+ \frac{f_s \cP_{-\,y}+\cP_{-\,ys}+\cP_{-\,s}
f_y}{4\cP_{-}}.
\lab{Atheta}
\ee
We will now show that
\be
\(rq \v_{\theta}\)_s =\(rq \v_{\theta=0}\)_s- \(\h f_y+g\)_{ys}
=
\h (f_{sy}+f_sf_y)_y+ \(\frac{f_s \cP_{-y}+\cP_{-ys}+\cP_{-s}
f_y}{2\cP_{-}}\)_y.
\lab{ch3}
\ee
Using equation \rf{rqzero}
one can easily show that equation \rf{ch3} holds if
the following relation
\[ - \( f_y+\cP_{-}\)_{s} = \frac{f_s \cP_{-\,y}+\cP_{-\,ys}+
\cP_{-\,s}f_y}{2\cP_{-}}
\]
is true. We note that the above relation can be rewritten as
\[
(\cP_{-}^2)_s+ 2 f _{ys} \cP_{-} +f_s \cP_{-\,y}+\cP_{-\,sy} +
\cP_{-\,s} f_y=0\, .
\]
The last equation is fully equivalent to the two-component Camassa--Holm
equation \rf{qsch} as can be seen by rewriting $Q$ from
relation \rf{Q-defg} as
$Q= -(\cP_{-}+f_y/2)^2-(\cP_{-}+f_y/2)_y$.
This completes the proof for relation \rf{ch3}.

It follows from \rf{r-der} and $ C=1/(16\b^2 B)- A^2/B$
that
\[ rq \v_{\theta} = \h \( {\bar f}_{yy}-\h {\bar f}_y^2-
\frac{{\bar f}_{sy}^2}{2 {\bar f}_s^2}
+\frac{1}{2 \b^2 {\bar f}_s^2}+ \frac{{\bar f}_{syy}}{{\bar f}_s}\).
\]
Thus, due to \rf{Atheta} and \rf{ch3} we have proved explicitly
that
\be
{\bar f}=f + \ln \( \tan \theta \( -
\frac{f_{y}}{2} +\frac{f_{ sy}}{2 f_{ s}}
+\frac{1}{2 \b f_{ s}} \)\) = \ft + \ln \cP_{-} (\ft)
\lab{back}
\ee
is a solution of a 2-component version of the Camassa--Holm equation.
Thus the transformation
\[
 f  \to  {\bar f}
\]
maps a solution $f$ of a 2-component version of the Camassa--Holm
equation to a dif\/ferent solution~${\bar f}$.
For example, let us consider, as in  \cite{Aratyn:2006vc},
the Camassa--Holm function:
\be
f(y,s)=\ln  \frac{a_1^{(1)}a_2^{(1)}z_1e^{\frac{s}{2z_1}+2yz_1}
+a_1^{(2)}a_2^{(2)}z_2e^{\frac{s}{2z_2}+2yz_2}
}
{
(z_2-z_1)a_1^{(2)}a_2^{(1)}
} ,
\lab{examp1}
\ee
where $a_i^{(j)}$, $i,j=1,2$ and $z_1$ and $z_2$ are constants.
The function $f$ solves equation \rf{CH-f} for $\b^2=1$.
Then, as an explicit calculation verif\/ies, the map $f \to {\bar f}$
with $\bar f$ given by expression~\rf{back} yields another solution
of  equation~\rf{CH-f} for $\b^2=1$ and $\theta\ne 0$.

For $\theta=\pi/2$ we have $r = \exp (-f)$ and comparing with the result for
$0<\theta <\pi/2$:
\be
r  = \cP_{+}  (\ft) e^{-\ft} ,
\lab{ref2}
\ee
we get a B\"{a}cklund transformation
\[
f \to  \ft-\ln \(  \cP_{+} (\ft) \)=
f_{\theta} - \ln \(
\frac{f_{\theta y}}{2} +\frac{f_{\theta  sy}}{2 f_{\theta s}}
+\frac{1}{2 \b f_{\theta s}} \)  .
\]
Additional B\"{a}cklund transformations can be obtained by
comparing expressions for $q$ and $r$ variables in terms
of~$f$ for the boundary values of~$\theta$.

We f\/irst turn our attention to the case of  $\theta=0$ for which
we have $ q = \exp (f)$ and
\be r =\h \( f_{yy}-\h f_y^2-
\frac{f_{sy}^2}{2 f_s^2}
+\frac{1}{2 \b^2 f_s^2}+ \frac{f_{syy}}{f_s}\)
e^{-f}
= \( \cP_{+}^2-\cP_{+}f_y + \cP_{+y}\) e^{-f}.
\lab{rth0}
\ee
From the AKNS equation \rf{CH-q} we see immediately that
$f = \ln q$ must satisfy the 2-component Camassa--Holm
equation \rf{CH-f}. Note, in addition, that the AKNS equation~\rf{CH-q}
is still valid if we replace $q$ by $r$ and therefore{\samepage
\[
f - \ln \( \cP_{+}^2-\cP_{+}f_y + \cP_{+y}\)
\]
must satisfy  the 2-component Camassa--Holm
equation \rf{CH-f} as well.}

Next, for $\theta=\pi/2$ we have $ r = \exp (-f)$ and
\be q =\h \(- f_{yy}-\h f_y^2-
\frac{f_{sy}^2}{2 f_s^2}
+\frac{1}{2 \b^2 f_s^2}+ \frac{f_{syy}}{f_s}\)
e^{f}
= \( \cP_{-}^2+\cP_{-}f_y + \cP_{-y}\)e^{f}.
\lab{qthpi2}
\ee
Comparing expressions for $q$ and $r$ we f\/ind
f\/ind that if $f$  is a solution of the 2-component Camassa--Holm
equation \rf{CH-f} then so is also
\[
f + \ln \( \cP_{-}^2+\cP_{-}f_y +\cP_{-y}\)  .
\]
To summarize we found the following B\"{a}cklund maps
\[
f \to \bigg\{ \begin{array}{l}
\ft\pm \ln \(  \cP_{\mp} (\ft) \), \qquad \ft=f +{\rm const},\vspace{2mm}\\
f \pm \ln \(\cP_{\mp}^2 \pm \cP_{\mp}f_y + \cP_{\mp y}\).
\end{array}
\]
The top row lists maps between $\theta=0,\pi/2$ cases and
$\theta$ within the interval $0 < \theta < \pi/2$ \cite{wu}.
The bottom row shows  new maps derived for
the $\theta=0$ and  $\pi/2$ cases only.

\section{Conclusions}
These notes describe an attempt to construct a general
and universal formalism which would realize possible connections between
the 2-component Camassa--Holm equation and AKNS hie\-rarchy extended by
a negative f\/low.

Construction yields gauge copies of an extended AKNS model
connected by a continuous parameter (angle) $\theta$
taking values in an interval
$ 0\le \theta \le \pi/2$.
Eliminating one of two components of the $sl(2)$ wave
function gives a second order non-linear
partial dif\/ferential equation for a single function $f$ of
the  two-component Camassa--Holm model.
Functions $f$ corresponding to dif\/ferent values of $\theta $
in an interior of interval $ 0\le \theta \le \pi/2$ dif\/fer
only by a trivial constant and fall into a~class considered in
\cite{CH2}. Two remaining and separate cases correspond
to  $\theta$ equal to $0$ and $\pi/2$ and agree with a structure
described in \cite{wu}.

\subsection*{Acknowledgements}

H.A. acknowledges partial support from Fapesp and IFT-UNESP
for their hospitality. JFG and AHZ thank CNPq for a partial support.

\LastPageEnding

\end{document}